\documentclass[aip,graphicx]{revtex4-1}
\usepackage[cp1251]{inputenc}
\usepackage[T2A]{fontenc}
\usepackage[english]{babel}
\usepackage{amssymb,latexsym,amsmath,amscd}
\usepackage{graphicx,color,framed}
\begin{document}
\selectlanguage{english}
\title{Statistical theory of fluids with a complex electric structure: Application to solutions of soft-core dipolar particles}
\author{\firstname{Yury A.} \surname{Budkov}}
\email[]{ybudkov@hse.ru}
\affiliation{Tikhonov Moscow Institute of Electronics and Mathematics, National Research University Higher School of Economics, Tallinskaya st. 34, 123458 Moscow, Russia}
\affiliation{G.A. Krestov Institute of Solution Chemistry of the Russian Academy of Sciences, Akademicheskaya st. 1, 153045 Ivanovo, Russia}
\begin{abstract}
Based on the thermodynamic perturbation theory (TPT) and the Random phase approximation (RPA), we present a statistical theory of solutions of electrically neutral soft molecules, every of which is modelled as a set of sites that interact with each other through the potentials, presented as the sum of the Coulomb potential and arbitrary soft-core potential. As an application of our formalism, we formulate a general statistical theory of solution of the soft-core dipolar particles. For the latter, we obtain a new analytical relation for the screening function. As a special case, we apply this theory to describing the phase behavior of a solution of the dipolar particles interacting with each other in addition to the electrostatic potential through the repulsive Gaussian potential -- Gaussian core dipolar model (GCDM). Using the obtained analytic expression for the total free energy of the GCDM, we obtain the liquid-liquid phase separation with an upper critical point. The developed formalism could be used as a general framework for the coarse-grained description of thermodynamic properties of solutions of macromolecules, such as proteins, betaines, polypeptides, etc.
\end{abstract}
\maketitle
\section{Introduction}
As is well known, in contrast to simple fluids whose molecules interact with each other through the potentials, always containing a short-range repulsive component (hard-core) \cite{Andersen1971,Andersen1972}, the intermolecular interaction in complex macromolecular fluids can be described by soft-core potentials \cite{Likos_2001,Molotilin2018}. The simplest soft-core potential, which can describe the effective interaction between the centers of mass of polymer coils in dilute and semi-dilute polymer solutions is the Gaussian core potential \cite{Stillinger1978}. So far, such a Gaussian core model (GCM) of fluid with the Gaussian intermolecular potential of interaction has been extensively studied for the bulk phase, as well as at the interfaces \cite{LouisGCM,Archer_1,Archer_2,Lang,Nogovitsin2007,Nogovitsin2006,Baeurle2006,Levin2018}. The most surprising result obtained by Louis et al. \cite{LouisGCM} is that thermodynamic properties of the GCM at sufficiently small energetic parameter of the repulsive interaction in a liquid state region can be described with good accuracy in a wide range of state parameters within the Random phase approximation (RPA). Comparing the analytical calculation with the Monte-Carlo simulation results, the authors showed the applicability of the RPA for the whole liquid state region of the system phase diagram. Thus, one can conclude that the availability of an accurate analytical equation of state for GCM allows us to use this model as a reference system for describing within the thermodynamic perturbation theory (TPT) more complex soft-core macromolecular fluids whose molecules carry additional microscopic quantities, such as multipole moments, polarizability, quadrupolarizability, {\sl etc}.

One can especially stress, among others, a number of macromolecules (such as proteins, polypeptides and betaines) that have some important applications and under certain conditions acquire a large dipole moment  ($\simeq 100-1000~D$) \cite{Canchi,Haran,Heldebrant2010,Kudaibergenov,Lowe,Felder_2007,Fedotova2016,Kumar_2009,Eiberwiser2015}. However, in order to describe theoretically the electrostatic interactions between such big dipolar macromolecules, it is incorrect to model them as the point-like dipoles (or even dipolar hard spheres) similar to polar molecules in the theory of simple dipolar fluids \cite{Onsager1936,Kirkwood1939,Hoye1976,Morozov2007,Levin1999,Nienhuis1972,Zhuang2018,
Abrashkin2007,Coalson1996,Budkov2015,BudkovJCP2016,BudkovEA2018,McEldrew2018} due to a very big dipole length ($\sim 1-20~nm$). Instead, it is necessary to consider a dipolar molecule as a set of charged centers (sites), separated by a fixed or fluctuating distance \cite{Budkov2018}. So far, only several analytical models of dipolar fluids taking into account the internal electric structure of dipolar molecules \cite{Chandler1977,Buyukdagli2013_JCP,Buyukdagli2013_PRE,
Blossey2014,Blossey2014_2,Budkov2018,BudkovJML2019,Gordievskaya2018,Martin2016} and only three of them \cite{Budkov2018,Gordievskaya2018,Martin2016} discuss possible applications to thermodynamical description of soft matter.

In paper \cite{Gordievskaya2018}, the authors proposed a statistical theory of conformational behavior of a dipolar polymer chain. In contrast to the previous theoretical models of solutions of dipolar polymer chains \cite{BudkovEPJE1,BudkovEPJE2,BudkovJCP2015,KolesnikovSoftMat2017,
Schiessel1998,Dua2014,Cherstvy2010}, modeling the electrostatic interaction of monomers as an interaction of point-like dipoles, the authors considered the monomeric units of dipolar chain as hard charged dumbbells \cite{Dussi2013}. Accounting properly for the conformational entropy, the excluded volume interactions of hard dumbbells and their electrostatic interactions, the authors achieved good agreement between the theoretical and MD simulation results for the conformational behavior of dipolar chain. The authors pointed out that their model might be used for qualitative describing the conformational behavior of polyzwitterions, such as betaines in polar solvents, folding of the proteins with the polar residues, weak polyelectrolytes and polyelectrolyte-based gels. In paper \cite{Martin2016}, the authors formulated a field-theoretic methodology for describing the thermodynamic behavior of polarizable soft matter. The model is based on fluid elements, referred to as beads, which can carry a net monopole charge at their center of mass and a permanent or induced dipole, described through a Drude-type distributed charge approach \cite{Martin2016}. Beads of different types can be mixed or linked into polymer chains with an arbitrary mechanism of chain flexibility \cite{Martin2016}. The authors claim that their general field theoretic model can be relevant for the description of a vast range of soft matter systems, such as polyelectrolytes, ionic liquids, polymerized ionic liquids, ionomers, block copolymers, among others. A nonlocal field-theoretic model has been recently proposed in our paper \cite{Budkov2018}, which is a generalization of formerly formulated a field-theoretical model \cite{Buyukdagli2013_PRE} of dipolar fluid with a fixed distance between the charged groups of dipolar particles. We have developed a model of dipolar particles as dimers of oppositely charged point-like polar groups, separated by a fluctuating distance, distributed in accordance with a certain probability distribution function $g(\bold{r})$. Using the proposed nonlocal field-theoretical model, we have derived a non-linear integro-differential equation with respect to the mean-field electrostatic potential, generalizing the Poisson-Boltzmann-Langevin equation for the point-like dipoles obtained in papers \cite{Abrashkin2007,Coalson1996} and generalized for the case of a polarizable solvent in \cite{Budkov2015}. We have used the obtained equation in its linearized form to derive expressions for the mean-field electrostatic potential of the point-like test ion and its solvation free energy in a salt-free solution, as well as in a solution in the presence of salt ions. As the main result of that paper, we obtained a general expression for the bulk electrostatic free energy of the ion-dipole mixture within the Random phase approximation (RPA). In addition, we pointed out that the developed theory could be applied to thermodynamic description of a solution of dipolar biomacromolecules and showed that nonlocal effects, related to the nonzero length of dipolar particles must be important for the aqueous solutions of proteins
with the dipole length $\approx 20~nm$ (dipole moment $\approx 1000~D$) that is quite accessible for the real protein aqueous solutions (see, for instance, \cite{Felder_2007}). We would like to note, however, that formulated theory could be applied to description of thermodynamic behavior of solutions of the dipolar macromolecules only at rather small concentrations, since it does not take into account the excluded volume interactions of the dipolar particles. In order to describe the behavior of thermodynamic functions of macromolecular solutions in a wide range of concentration, it is necessary to take into account not only the electrostatic interactions, but also the soft-core repulsive interactions. In papers \cite{Buyukdagli2011,Buyukdagli2012} the authors formulated a field-theoretical model of electrolyte solution whose ions interact with each other via the Coulomb and soft-core (Yukawa) repulsive potentials. The authors applied this theory to description of ionic liquid in the bulk, as well as at the neutral dielectric interface with account for the finite size of ions (modelled by the soft-core repulsive potential) within the variational method. However, theoretical description of the thermodynamic properties of soft matter systems with more complex internal electric structure than that of the ions (dipolar and quadrupolar particles, for instance) requires formulating a model of soft-core molecules that are modelled as a set of spatially correlated sites interacting with each other through the Coulomb and soft-core potentials. In present paper, we will formulate such a field theory, based on TPT and RPA for description of thermodynamic properties of solutions of soft-core molecules with a complex electric structure. As an illustration, we will show how the developed formalism can be applied to description of the phase behavior of the solution of soft-core dipolar particles.

The rest of the article is organized as follows. In sec. II, we propose a general field-theoretic formalism based on TPT and RPA for a theoretical description of thermodynamic properties of solutions of soft-core particles with an arbitrary internal electric structure. In Sec. III, we apply the developed formalism to the formulation of the general theory of soft-core dipolar particles, generalizing the previously formulated theory \cite{Budkov2018}. Sec. IV considers the simplified Gaussian-core dipolar model (GCDM), which is a particular case of the general model, discussed in Sec. III. Sec. V presents some numerical results regarding the phase behavior of solution of Gaussian core dipolar particles and Sec. VI contains some concluding remarks and further prospects.

\section{General formalism}
In this section, based on the thermodynamic pertubation theory (TPT) and the Random phase approximation (RPA), we will formulate a general field theoretical approach to the description of solutions of electrically neutral soft particles with a complex internal electric structure.
\subsection{Derivation of excess free energy of fluids with complex electric structure}
Let us consider a solution of $N$ electrically neutral molecules which we will model as a set of $l$ charged sites with charges $q_{\alpha}$ ($\alpha=1,..,l$), so that $\sum\limits_{\alpha=1}^{l}q_{\alpha}=0$. Let us assume that in each molecule, the positions $\bold{r}^{(\alpha)}$ of sites are random variables, described by the probability distribution function $g(\bold{r}^{(1)},..,\bold{r}^{(l)})$, which in general case is determined by the quantum nature of the molecule and have to satisfy the following normalization condition
\begin{equation}
\int..\int d\bold{r}^{(\alpha_{1})}..d\bold{r}^{(\alpha_{l-1})}g(\bold{r}^{(1)},..,\bold{r}^{(l)})=1,
\end{equation}
where $\alpha_{i}\neq\alpha_{k}$ for all $i\neq k$.
As in our previous paper \cite{Budkov2018}, we consider the case of an implicit solvent, modelling it as a continuous dielectric medium with the constant dielectric permittivity $\varepsilon$. Let us also assume that apart from the Coulomb interactions between sites there are also some "volume" non-electrostatic interactions, described by the $l\times l$ matrix of soft-core potentials $V_{\alpha\gamma}(\bold{r})$.

Therefore, bearing in mind all the above model assumptions, we can write the configuration integral of the solution as follows
\begin{equation}
Q=\int d\Gamma \exp\left[-\beta H_{0}-\beta H_{el}\right],
\end{equation}
where the integration measure is
\begin{equation}
\int d\Gamma(\cdot)=\int..\int\prod\limits_{i=1}^{N}\frac{d\bold{r}_{i}^{(1)}..d\bold{r}_{i}^{(l)}}{V}g(\bold{r}_{i}^{(1)},..,\bold{r}_{i}^{(l)})(\cdot),
\end{equation}
$\beta=(k_{B}T)^{-1}$ is the inverse thermal energy. The Hamiltonian of non-electrostatic interactions is
\begin{equation}
H_{0}=\frac{1}{2}\int d\bold{r}\int d\bold{r}^{\prime}\sum\limits_{\alpha,\gamma}\hat{n}_{\alpha}(\bold{r})V_{\alpha\gamma}(\bold{r}-\bold{r}^{\prime})\hat{n}_{\gamma}(\bold{r}^{\prime})=\frac{1}{2}\left(\hat{n}V\hat{n}\right),
\end{equation}
where $\hat{n}_{\alpha}(\bold{r})=\sum\limits_{i=1}^{N}\delta\left(\bold{r}-\bold{r}_{i}^{(\alpha)}\right)$ is the microscopic concentration of the sites of $\alpha^{th}$ type and $V_{\alpha\gamma}(\bold{r}-\bold{r}^{\prime})$ is the matrix of the non-electrostatic interaction potentials between the sites of molecules. The Hamiltonian of the electrostatic interactions is
\begin{equation}
H_{el}=\frac{1}{2}\int d\bold{r}\int d\bold{r}^{\prime}\hat{\rho}(\bold{r})G_{0}(\bold{r}-\bold{r}^{\prime})\hat{\rho}(\bold{r}^{\prime})=\frac{1}{2}\left(\hat{\rho} G_{0}\hat{\rho}\right),
\end{equation}
where $\hat{\rho}(\bold{r})=\sum\limits_{\alpha=1}^{l}q_{\alpha}\hat{n}_{\alpha}(\bold{r})$ is the microscopic charge density of molecules satisfying the electroneutrality condition, i.e. $\int d\bold{r}\hat{\rho}(\bold{r})=0$; $G_{0}(\bold{r}-\bold{r}^{\prime})=1/(4\pi\varepsilon\varepsilon_{0}|\bold{r}-\bold{r}^{\prime}|)$ is the Green function of the Poisson equation.

Further, using the standard Hubbard-Stratonovich transformation
\begin{equation}
\exp\left[-\frac{\beta}{2}(\hat\rho G_0\hat\rho)\right]=\int\frac{\mathcal{D}\varphi}{C}\exp\left[-\frac{\beta}{2}(\varphi G_0^{-1}\varphi)+i\beta(\hat\rho\varphi)\right],
\end{equation}
we arrive at the following functional representation of the configuration integral
\begin{equation}
\label{conf_int}
Q=\int\frac{\mathcal{D}\varphi}{C}\exp\left[-\frac{\beta}{2}(\varphi G_0^{-1}\varphi)\right]\int d\Gamma \exp\left[-\beta H_{0}\right]\exp\left[i\beta (\hat{\rho}\varphi)\right],
\end{equation}
where the following short-hand notations
$$(\varphi G_0^{-1}\varphi)=\int d\bold r\int d\bold r'\varphi(\bold r)G_0^{-1}(\bold r,\bold r')\varphi(\bold r'),~~(f\varphi)=\int d\bold r f(\bold r)\varphi(\bold r)$$
and the normalization constant of the Gaussian measure
$$C=\int \mathcal{D}\varphi\exp\left[-\frac{\beta}{2}(\varphi G_0^{-1}\varphi)\right]$$
have been introduced. Considering a solution of molecules without Coulomb interactions as a reference system, one can write configuration integral (\ref{conf_int}) in the standard TPT form:
\begin{equation}
\label{conf_int_TPT}
Q=Q_{0}\int\frac{\mathcal{D}\varphi}{C}\exp\left[-\frac{\beta}{2}(\varphi G_0^{-1}\varphi)\right]\langle\exp\left[i\beta (\hat{\rho}\varphi)\right]\rangle_{0},
\end{equation}
where
\begin{equation}
Q_{0}=\int d\Gamma \exp\left[-\beta H_{0}\right]
\end{equation}
is the configuration integral of the reference system, which will be calculated below and the symbol
\begin{equation}
\langle(\cdot)\rangle_{0}=\frac{1}{Q_0}\int d\Gamma \exp\left[-\beta H_{0}\right](\cdot)
\end{equation}
denotes the average over the statistics of the reference system. Using the Random phase approximation (RPA), i.e. truncating the cumulant expansion \cite{Kubo} in the integrand exponent by the second term, we arrive at
\begin{equation}
\label{conf_int_RPA}
Q\approx Q_{0}\int\frac{\mathcal{D}\varphi}{C}\exp\left[-\frac{\beta}{2}(\varphi G_0^{-1}\varphi)-\frac{\beta^2}{2}(\varphi S_{c}\varphi)\right],
\end{equation}
where
\begin{equation}
S_c(\bold{r}-\bold{r}^{\prime})=\langle \hat{\rho}(\bold{r})\hat{\rho}(\bold{r}^{\prime})\rangle_0 =
\sum\limits_{\alpha,\gamma}q_{\alpha}q_{\gamma}S_{\alpha\gamma}(\bold{r}-\bold{r}^{\prime})
\end{equation}
is the correlator of microscopic charge density with the structure matrix $S_{\alpha\gamma}(\bold{r}-\bold{r}^{\prime})$ of the reference system, whose elements can be calculated as follows
\begin{equation}
S_{\alpha\gamma}(\bold{r}-\bold{r}^{\prime})=\langle \delta \hat{n}_{\alpha}(\bold{r})\delta \hat{n}_{\gamma}(\bold{r}^{\prime})\rangle_0,
\end{equation}
where the local fluctuations of microscopic concentrations of the sites $\delta\hat{n}_{\alpha}(\bold{r})=\hat{n}_{\alpha}(\bold{r})-\langle\hat{n}_{\alpha}(\bold{r})\rangle_0 $ are introduced.

Therefore in the RPA, the configuration integral takes the form \cite{Podgornik1989}
\begin{equation}
\label{conf_int_RPA_2}
Q\approx Q_{0}\int\frac{\mathcal{D}\varphi}{C}\exp\left[-\frac{\beta}{2}(\varphi G^{-1}\varphi)\right]=Q_0 \exp\left[\frac{1}{2}tr\left(\ln{G}-\ln{G_0}\right)\right],
\end{equation}
where the symbol $tr(A)$ denotes the trace of integral operator, so that for the kernel $A(\bold{r},\bold{r}^{\prime})$ we have the following relation
\begin{equation}
tr(A)=\int d\bold{r}A(\bold{r},\bold{r})
\end{equation}
and
\begin{equation}
\label{Green}
G^{-1}(\bold{r}-\bold{r}^{\prime})=G^{-1}_0(\bold{r}-\bold{r}^{\prime})+\beta S_{c}(\bold{r}-\bold{r}^{\prime})
\end{equation}
is the renormalized reciprocal Green function of the solution medium in RPA. We would like to note that trace of operator $A$ with a translation invariant kernel $A(\bold{r},\bold{r}^{\prime})=A(\bold{r}-\bold{r}^{\prime})$ can be calculated as
\begin{equation}
tr(A)=V\int\frac{d\bold{k}}{(2\pi)^3}\tilde{A}(\bold{k}),
\end{equation}
where $V$ is the total volume of system and
\begin{equation}
\tilde{A}(\bold{k})=\int d\bold{r}A(\bold{r})e^{-i\bold{k}\bold{r}}
\end{equation}
is the Fourier-image of function $A(\bold{r})$.

Therefore, taking into account the expression (\ref{Green}) and definitions of trace and subtracting the electrostatic self-energy of molecules from the final expression, we arrive at the total excess free energy
\begin{equation}
\label{ex_free_en_RPA}
F_{ex}=F_{0}+\frac{Vk_{B}T}{2}\int\frac{d\bold{k}}{(2\pi)^3}\left(\ln\left(1+\frac{\varkappa^2(\bold{k})}{k^2}\right)-\frac{\varkappa^2(\bold{k})}{k^2}\right),
\end{equation}
where
\begin{equation}
\label{screen_func}
\varkappa^2(\bold{k})=\frac{1}{k_{B}T\varepsilon\varepsilon_0}\sum\limits_{\alpha,\gamma}q_{\alpha}q_{\gamma}S_{\alpha\gamma}(\bold{k})
\end{equation}
is the screening function \cite{Borue1988,Brilliantov1993,Lue2006,Victorov2016,Khokhlov1982} and $S_{\alpha\gamma}(\bold{k})$ are the structure factors of the reference system. It should be noted that equation (\ref{screen_func}) has been obtained first in work \cite{Borue1988} within the different approach.

\subsection{Calculation of free energy and structure factors of the reference system within RPA}
In order to calculate the excess free energy (\ref{ex_free_en_RPA}) of solution, it is necessary to calculate the excess free energy $F_0$ and the structure factors $S_{\alpha\gamma}(\bold{k})$ of the reference system. For this purpose, we determine the following auxiliary configuration integral of the reference system
\begin{equation}
\label{Q_0_psi}
Q_0[\Psi]=\int d\Gamma \exp\left[-\frac{\beta}{2}\left(\hat{n}\bold{V}\hat{n}\right)-\beta\left(\delta\hat{n}\Psi\right)\right],
\end{equation}
where $\left(\delta\hat{n}\Psi\right)=\int d\bold{r}\sum\limits_{\alpha}\delta \hat{n}_{\alpha}(\bold{r})\Psi_{\alpha}(\bold{r})$ with the auxiliary external potentials $\Psi_{\alpha}(\bold{r})$. The structure factors can be calculated as
\begin{equation}
\label{func_deriv}
S_{\alpha\gamma}(\bold{r}-\bold{r}^{\prime})=
\frac{1}{\beta^2}\frac{\delta^2}{\delta\Psi_{\alpha}(\bold{r})\delta\Psi_{\gamma}(\bold{r}^{\prime})}\frac{Q_0[\Psi]}{Q_0[0]}\biggr\rvert_{\Psi=0}.
\end{equation}

Further, using the Hubbard-Stratonovich transformation
\begin{equation}
e^{-\frac{\beta}{2}\left(\hat{n}\bold{V}\hat{n}\right)}=\int\frac{\mathcal{D}\Phi}{C_{V}}e^{-\frac{\beta}{2}\left(\Phi \bold{V}^{-1}\Phi\right)+i\beta (\hat{n}\Phi)},
\end{equation}
where
$$\left(\Phi \bold{V}^{-1}\Phi\right)=\int d\bold{r}\int d\bold{r}^{\prime}\sum\limits_{\alpha,\gamma}\Phi_{\alpha}(\bold{r})V^{-1}_{\alpha\gamma}(\bold{r}-\bold{r}^{\prime})\Phi_{\gamma}(\bold{r}^{\prime}),~~(\hat{n}\Phi)=\int d\bold{r}\sum\limits_{\alpha}\hat{n}_{\alpha}(\bold{r})\Phi_{\alpha}(\bold{r}),$$
and $C_{V}=\int\mathcal D\Phi e^{-\frac{\beta}{2}\left(\Phi \bold{V}^{-1}\Phi\right)}$ is the renormalization constant of the Gaussian measure, we obtain the following representation of (\ref{Q_0_psi}):
\begin{equation}
\label{gen_func}
Q_{0}[\Psi]=\int\frac{\mathcal{D}\Phi}{C_{V}}e^{-\frac{\beta}{2}\left(\Phi \bold{V}^{-1}\Phi\right)}\int d\Gamma e^{i\beta (\hat{n}\Phi)-\beta (\delta\hat{n}\Psi)},
\end{equation}
where the reciprocal matrix operator $\bold{V}^{-1}$ is determined by
\begin{equation}
\int d\bold{r}^{\prime\prime}\sum\limits_{\lambda}^{}
V_{\alpha\lambda}(\bold{r}-\bold{r}^{\prime\prime})V^{-1}_{\lambda\gamma}(\bold{r}^{\prime\prime}-\bold{r}^{\prime})=
\delta_{\alpha\gamma}\delta(\bold{r}-\bold{r}^{\prime}).
\end{equation}

Further, rewriting the fluctuating field variable as $\Phi(\bold{r})=\Phi_{0}+\Lambda(\bold{r})$, where $\Phi_{0}=\frac{1}{V}\int d\bold{r}\Phi(\bold{r})$, and performing the integration over the constant field $\Phi_{0}$ in (\ref{gen_func}), we obtain
\begin{equation}
Q_{0}[\Psi]=Q_{0}^{(MF)}e^{\beta (n\Psi)}\int\frac{\mathcal{D}\Lambda}{C_{V}}e^{-\frac{\beta}{2}\left(\Lambda \bold{V}^{-1}\Lambda\right)}\int d\Gamma e^{i\beta (\hat{n}[\Lambda+i\Psi])},
\end{equation}
where
\begin{equation}
Q_{0}^{(MF)}=\exp\left[-\frac{\beta}{2}\left(n \bold{V} n\right)\right]
\end{equation}
is the mean-field approximation for the configuration integral of the reference system with the average concentrations $n_{\alpha}(\bold{r})=\langle \hat{n}_{\alpha}(\bold{r})\rangle_0$ of the charged sites. Now, performing the following shift $\Lambda\rightarrow \Lambda -i \Psi$, we obtain
\begin{equation}
Q_{0}[\Psi]=Q_{0}^{(MF)}e^{\beta (n\Psi)+\frac{\beta}{2}\left(\Psi \bold{V}^{-1}\Psi\right)}\int\frac{\mathcal{D}\Lambda}{C_{V}}e^{-\frac{\beta}{2}\left(\Lambda \bold{V}^{-1}\Lambda\right)+i\beta\left(\Lambda \bold{V}^{-1}\Psi\right)}Q^{N}[\Lambda],
\end{equation}
where
\begin{equation}
Q[\Lambda]=\int..\int\frac{d\bold{r}^{(1)}..d\bold{r}^{(l)}}{V}g(\bold{r}^{(1)},..,\bold{r}^{(l)})\exp\left[i\beta\sum\limits_{\alpha=1}^{l}\Lambda_{\alpha}(\bold{r}^{(\alpha)})\right]
\end{equation}
is the one-particle partition function. In the thermodynamic limit $N\to \infty$, $V\to \infty$, $N/V\to n$, we obtain \cite{Efimov1996,Budkov2018}
\begin{equation}
Q_{0}[\Psi]=Q_{0}^{(MF)}e^{\beta (n\Psi)+\frac{\beta}{2}\left(\Psi \bold{V}^{-1}\Psi\right)}\int\frac{\mathcal{D}\Lambda}{C_{V}}e^{-\frac{\beta}{2}\left(\Lambda \bold{V}^{-1}\Lambda\right)+i\beta\left(\Lambda \bold{V}^{-1}\Psi\right)+W[\Lambda]},
\end{equation}
where we have introduced the following functional
\begin{equation}
W[\Lambda]=n\int..\int d\bold{r}^{(1)}..d\bold{r}^{(l)} g(\bold{r}^{(1)},..,\bold{r}^{(l)})\left(\exp\left[i\beta\sum\limits_{\alpha=1}^{l}\Lambda_{\alpha}(\bold{r}^{(\alpha)})\right]-1\right).
\end{equation}
Taking into account that $\int d\bold{r}\Lambda(\bold{r})=0$, expanding the functional $W[\Lambda]$ in the power-law series and truncating it by the quadratic term, we obtain
\begin{equation}
\nonumber
Q_{0}[\Psi]\approx Q_{0}^{(MF)}e^{\beta (n\Psi)+\frac{\beta}{2}\left(\Psi \bold{V}^{-1}\Psi\right)}\int\frac{\mathcal{D}\Lambda}{C_{V}}e^{-\frac{\beta}{2}\left(\Lambda \bold{D}^{-1}\Lambda\right)+i\beta\left(\Lambda \bold{V}^{-1}\Psi\right)}
\end{equation}
\begin{equation}
=Q_{0}^{(RPA)}e^{\beta (n\Psi)+\frac{\beta}{2}\left(\Psi \bold{V}^{-1}\Psi\right)-\frac{\beta}{2}\left(\Psi \bold{V}^{-1}\bold{D}\bold{V}^{-1}\Psi\right)}
\end{equation}
where
\begin{equation}
Q_{0}^{(RPA)}=Q_{0}^{(MF)}\exp\left[\frac{1}{2}Tr\left(\ln \bold{D} - \ln \bold{V}\right) \right]
\end{equation}
is the reference system configuration integral in the RPA and operator $\bold{D}$ is determined by the relation
\begin{equation}
D^{-1}_{\alpha\gamma}(\bold{r}-\bold{r}^{\prime})=V^{-1}_{\alpha\gamma}(\bold{r}-\bold{r}^{\prime})+\beta W_{\alpha\gamma}(\bold{r}-\bold{r}^{\prime}),
\end{equation}
where the operator $\bold{W}$ has the following form
\begin{equation}
W_{\alpha\gamma}(\bold{r}-\bold{r}^{\prime})=n\left(\delta_{\alpha\gamma}\delta(\bold{r}-\bold{r}^{\prime})+
(1-\delta_{\alpha\gamma})g_{\alpha\gamma}(\bold{r}-\bold{r}^{\prime})\right)
\end{equation}
with the structure functions
\begin{equation}
\nonumber
g_{\alpha\gamma}(\bold{r}-\bold{r}^{\prime})=\int..\int d\bold{r}^{(1)}..d\bold{r}^{(\alpha-1)}d\bold{r}^{(\alpha+1)}..
d\bold{r}^{(\gamma-1)}d\bold{r}^{(\gamma+1)}..d\bold{r}^{(l)}\times
\end{equation}
\begin{equation}
\times g(\bold{r}^{(1)},..,\bold{r}^{(\alpha-1)},\bold{r},\bold{r}^{(\alpha+1)},..,\bold{r}^{(\gamma-1)},\bold{r}^{\prime},\bold{r}^{(\gamma+1)},..,\bold{r}^{(l)})
\end{equation}
which describe nothing more than the probability distribution functions of the distance between $\alpha^{th}$ and $\gamma^{th}$ sites. The symbol $Tr(..)$ denotes the trace of integral matrix operator in accordance with the following definition
\begin{equation}
\label{Tr}
Tr(\bold{A})=\int d\bold{r}\sum\limits_{\alpha}^{}A_{\alpha\alpha}(\bold{r},\bold{r}).
\end{equation}

Using relation (\ref{func_deriv}), we obtain the following result for the structure factor
\begin{equation}
\bold{S}=\frac{1}{\beta}\left(\bold{V}^{-1}-\bold{V}^{-1}\bold{D}\bold{V}^{-1}\right),
\end{equation}
which after some simple transformations can be rewritten as follows
\begin{equation}
S_{\alpha\gamma}^{-1}(\bold{r}-\bold{r}^{\prime})=W_{\alpha\gamma}^{-1}(\bold{r}-\bold{r}^{\prime})+\beta V_{\alpha\gamma}(\bold{r}-\bold{r}^{\prime})
\end{equation}
or in the Fourier-representation
\begin{equation}
\label{_fact}
S_{\alpha\gamma}^{-1}(\bold{k})=W_{\alpha\gamma}^{-1}(\bold{k})+\beta \tilde{V}_{\alpha\gamma}(\bold{k}).
\end{equation}

Therefore, using the defintion (\ref{Tr}) of the trace of integral matrix operator and taking into account that $$\frac{\det{\tilde{D}(\bold{k})}}{\det{\tilde{V}(\bold{k})}}=\frac{\det{S(\bold{k})}}{\det{W(\bold{k})}},$$ we obtain the free energy of the reference system in the RPA as follows
\begin{equation}
\label{F0}
F_0 = -k_{B}T\ln{Q_{0}^{(RPA)}}=\frac{V}{2}n^2\sum\limits_{\alpha,\gamma}\tilde{V}_{\alpha\gamma}(0)-
\frac{Vk_{B}T}{2}\int\frac{d\bold{k}}{(2\pi)^3}\ln\frac{\det{S(\bold{k})}}{\det{W(\bold{k})}}.
\end{equation}
The first term in the right hand side of eq. (\ref{F0}) determines the mean-field approximation for the contribution of excluded volume interactions to the total free energy, whereas the second term -- contribution of the Gaussian fluctuations of the potential energy near its average value.

\section{General theory of soft dipolar particles}
Now, we turn to the application of our general theory to describing dipolar particles. We assume that each dipolar particle is a dimer of the oppositely charged sites with charges $\pm q$, separated by the fluctuating distance. As in our earlier paper \cite{Budkov2018}, we introduce the probability distribution function $g(\bold{r})$ of the distance between the charged sites. The relation for the structure factor of the dipolar particles solution in the RPA takes the following form
\begin{equation}
\label{eq:matrix1}
\hat{S}^{-1}\left(\bold{k}\right)=\hat{W}^{-1}\left(\bold{k}\right)+\beta \tilde{V}(\bold{k})=
\begin{pmatrix}
\frac{1}{n\left(1-g^2(\bold{k})\right)}+\beta V_{11}(\bold{k}) & -\frac{g(\bold{k})}{n\left(1-g^2(\bold{k})\right)}+\beta V_{12}(\bold{k})\\
-\frac{g(\bold{k})}{n\left(1-g^2(\bold{k})\right)}+\beta V_{12}(\bold{k}) & \frac{1}{n\left(1-g^2(\bold{k})\right)}+\beta V_{22}(\bold{k})\\
\end{pmatrix}
,
\end{equation}
where
\begin{equation}
\label{eq:matrix1}
\tilde{V}\left(\bold{k}\right)=
\begin{pmatrix}
\tilde{V}_{11}(\bold{k}) & \tilde{V}_{12}(\bold{k})\\
\tilde{V}_{12}(\bold{k}) & \tilde{V}_{22}(\bold{k})\\
\end{pmatrix}
\end{equation}
is the matrix of Fourier-images of non-electrostatic interaction potentials between sites of dipolar particles, whereas the structure matrix has the following form
\begin{equation}
\label{eq:matrix1}
W\left(\bold{k}\right)=
\begin{pmatrix}
n & n g(\bold{k})\\
n g(\bold{k}) & n\\
\end{pmatrix},
\end{equation}
where
\begin{equation}
g(\bold{k})=\int d\bold{r}g(\bold{r})e^{-i\bold{k}\bold{r}}
\end{equation}
is the characteristic function corresponding to the probability distribution function $g(\bold{r})$.

Further, using relations (\ref{screen_func}) and (\ref{eq:matrix1}), we obtain the following new analytical expression for the screening function for the solution of soft-core dipolar particles
\begin{equation}
\label{screen_func_2}
\varkappa^2(\bold{k})=\frac{\kappa_{D}^{2}\left(1-g(\bold{k})\right)\left(1+\frac{\beta n}{2}\left(\tilde{V}_{11}(\bold{k})+
\tilde{V}_{22}(\bold{k})+2\tilde{V}_{12}(\bold{k})\right)\left(1+g(\bold{k})\right)\right)}{1+\beta n\left(\tilde{V}_{11}(\bold{k})+
\tilde{V}_{22}(\bold{k})+2\tilde{V}_{12}(\bold{k})g(\bold{k})\right)+\beta^2n^2\left(1-g^2(\bold{k})\right)\left(\tilde{V}_{11}(\bold{k})\tilde{V}_{22}(\bold{k})-\tilde{V}_{12}^2(\bold{k})\right)},
\end{equation}
where $\kappa_{D}^{2}=r_{D}^{-2}= 2q^2n/(\varepsilon\varepsilon_0 k_{B}T)$ is the square of the inverse Debye radius $r_D$, attributed to the charged sites of dipolar particles. Note that in the absence of the excluded volume interactions ($\tilde{V}_{\alpha\gamma}(\bold{k})=0$) between the sites, the expression (\ref{screen_func_2}) will transform into that has been obtained earlier in \cite{Budkov2018}.

\section{Theory of Gaussian core dipolar particles}
In order to obtain the analytical results for the excess free energy of a dipolar particles solution, we will simplify the general model, considering only the case of Gaussian core dipolar model (GCDM). Namely, let us consider a reference system with $\tilde{V}_{11}(\bold{k})=\tilde{V}(\bold{k})>0$ and $\tilde{V}_{22}(\bold{k})=\tilde{V}_{12}(\bold{k})=0$. Essentially, such a reference system describes a set of soft particles with a point-like $"$fantom$"$ particle, grafted to their center of mass. It should be noted that this theoretical model could be applied for coarse-grained description of the macromolecules of dipolar polypeptides or betaines, dissolved in water. Moreover, such theoretical model could be applied to coarse-grained description of zwitterionic liquids whose molecules are usually comprised of the covalently bounded big organic cation and small anion \cite{Heldebrant2010}.

Thus, in that case we obtain the following relation for the screening function:
\begin{equation}
\varkappa^2(\bold{k})=\kappa_{D}^{2}\left(1-g(\bold{k})\right)Q(\bold{k}),
\end{equation}
where
\begin{equation}
Q(\bold{k})=1-\frac{1}{2}\frac{\beta n\tilde{V}(\bold{k})}{1+\beta n \tilde{V}(\bold{k})}\left(1-g(\bold{k})\right)
\end{equation}
is the auxiliary function. The electrostatic free energy can be written as follows
\begin{equation}
\label{Fcorr3}
F_{el} =\frac{Vk_{B}T}{2}\int\frac{d\bold{k}}{(2\pi)^3}\left(\ln\left(1+\frac{\kappa_{D}^2}{k^2}\left(1-g(\bold{k})\right)Q(\bold{k})\right)-
\frac{\kappa_{D}^2}{k^2}\left(1-g(\bold{k})\right)Q(\bold{k})\right).
\end{equation}
As in our previous paper \cite{Budkov2018}, to describe a distribution function $g(\bold{r})$ of the distance between charged groups of dipolar particles, we use the Yukawa-type distribution function
\begin{equation}
g(\bold{r})=\frac{3}{2\pi l^2r}\exp\left[-\sqrt{6}r/l\right],
\end{equation}
which determines the following characteristic function
\begin{equation}
\label{char}
g(\bold{k})=\frac{1}{1+\frac{k^2l^2}{6}}.
\end{equation}
Using the expression (\ref{char}), following 'mean-field' approximation
\begin{equation}
Q(\bold{k})\approx 1-\frac{1}{2}\frac{\beta n \tilde{V}(0)}{1+\beta n \tilde{V}(0)}\left(1-g(\bold{k})\right)=1-\frac{\alpha}{2}\left(1-g(\bold{k})\right),
\end{equation}
and taking integral (\ref{Fcorr3}), we arrive at the new analytical expression for the electrostatic free energy of solution of the dipolar particles
\begin{equation}
\label{Fcorr4}
F_{el}=-\frac{V k_{B}T}{l^3}\theta(y,\alpha),
\end{equation}
where
\begin{equation}
\nonumber
\theta(y,\alpha)=\left(1-\frac{3\alpha}{4}\right)\sigma(y)-\frac{3\sqrt{6}\left((\alpha+4)(y^2+2y+y\sqrt{1+y})+8(1+\sqrt{1+y})\right)}{8\pi\left(1+\sqrt{1+y}\right)}+
\end{equation}
\begin{equation}
\label{sigma2}
+\frac{3\sqrt{6}\exp\left[\frac{\alpha y}{4(1+\sqrt{1+y})^2}\right]\left(y\left(4+\sqrt{1+y}(\alpha+2)\right)+2y^2+4(1+\sqrt{1+y})\right)}{4\pi (1+\sqrt{1+y})}.
\end{equation}
Here, $\alpha=\beta n \tilde{V}(0)/(1+\beta n \tilde{V}(0))<1$, $y=l^2/6r_D^2$ is the strength of dipole-dipole interactions and the auxiliary function \cite{Budkov2018}
\begin{equation}
\sigma(y)=\frac{\sqrt{6}}{4\pi}\left(2(1+y)^{3/2}-2-3y\right)
\end{equation}
has been introduced.

As in the case of point-like dipolar particles \cite{Budkov2018}, one can obtain two limiting regimes for the electrostatic free energy
\begin{equation}
\label{eq:pot}
\frac{F_{el}}{Vk_BT}=
\begin{cases}
-\frac{\sqrt{6}e^4 l n^2}{48\pi \left(\varepsilon\varepsilon_{0}k_{B}T\right)^2}\left(1-\frac{3\alpha}{4}+\frac{5\alpha^2}{32}\right), y\ll 1\,\\
-\frac{1}{12\pi {r_D}^3}\left(1-\frac{3}{2}\alpha-3\left(1-e^{\alpha/4}\right)\right),y\gg 1.
\end{cases}
\end{equation}
As in the theory without the repulsive interactions \cite{Budkov2018} between dipolar particles, the first regime ($r_D\gg l$) corresponds to the situation, when the electrostatic interactions of charged groups manifest themself as the Keesom interactions \cite{Budkov2018} of dipoles, separated by rather big distances; the second regime ($r_D\ll l$) corresponds to the case when the charged groups of the dipolar particles do not "feel" connectivity anymore and, thus, behave as the freely moving ions in regular electrolyte solution, participating in the screening of electrostatic interactions. As is seen from relations (\ref{eq:pot}), in both limiting regimes, the occurrence of an additional repulsive interaction between dipolar particles results in a decrease in the absolute value of the electrostatic free energy with respect to the case of dipolar particles without excluded volume interactions \cite{Budkov2018}. Such a behavior can be explained by the fact that occurrence of the finite volume of the dipolar particles prevents their mutual approaching to the small distances that leads to decrease in the electrostatic free energy in its absolute value.

Using the general expression (\ref{F0}), we obtain the following simple relation for the excess free energy of the reference system, which has the form of the free energy of a one-component fluid with a soft central potential of the intermolecular interaction with a positive Fourier-image \cite{Zakharov1999,Zubarev1954,Holowko_book}
\begin{equation}
F_{0}=\frac{Vn^2\tilde{V}(0)}{2}+\frac{Vk_{B}T}{2}\int\frac{d\bold{k}}{(2\pi)^3}\left(\ln\left(1+\beta n\tilde{V}(\bold{k})\right)-\beta n\tilde{V}(\bold{k})\right),
\end{equation}
from which we have subtracted the self-energy of molecules. Choosing the GCM \cite{LouisGCM} as a reference system, i.e. using the Gaussian potential for modeling the non-electrostatic interactions of dipolar particles
\begin{equation}
V(r)=V_0 \exp\left[-{r^2}/{R^2}\right]
\end{equation}
and taking into account that $\tilde{V}(\bold{k})=\pi^{3/2}R^3 V_0 \exp\left[-k^2R^2/4\right]$, we obtain the following expression for the reference system free energy
\begin{equation}
\label{Gaussian}
\frac{F_0}{Nk_{B}T} = \frac{\xi}{2}\left(1-\beta V_0 \mathcal{F}(\xi)\right),
\end{equation}
where $\xi=\pi^{3/2}\beta nV_0 R^3$, $R$ is the effective size of the Gaussian core and $V_0$ is the energetic parameter, showing the strength of the repulsive non-electrostatic interactions;
the auxiliary function
\begin{equation}
\mathcal{F}(\xi)=\frac{4}{3\sqrt{\pi}}\int\limits_{0}^{1}\frac{dt(-\ln{t})^{3/2}t}{1+\xi t}
\end{equation}
is introduced, which can be expressed via the polylogarithm function. We would like to note that the free energy (\ref{Gaussian}) describes the thermodynamic properties of the GCM in the liquid-state region with good accuracy at sufficiently small interaction parameter $V_0$ \cite{LouisGCM}.

\section{Numerical results and discussions}
Before turning to numerical calculations, we introduce the following set of dimensionless parameters: $\tilde{n}=nR^3$, $\tilde{T}=4\pi\varepsilon\varepsilon_0 R k_{B}T/q^2$, $\tilde{V}_0=4\pi\varepsilon\varepsilon_0 R V_0/q^2$, $\tilde{l}=l/R$. Due to the fact that for the real dipolar macromolecules (such as betaines or polypeptides) the effective dipole length $l$ is order of the effective size of core $R$, we assume in further that $\tilde{l}=1$. In what follows we will consider only sufficiently small values of energetic parameter $V_0$ of Gaussian potential to satisfy the applicability condition of the RPA for the GCM \cite{LouisGCM,Likos_2001}.

\begin{figure}[h!]
\center{\includegraphics[width=0.8\linewidth]{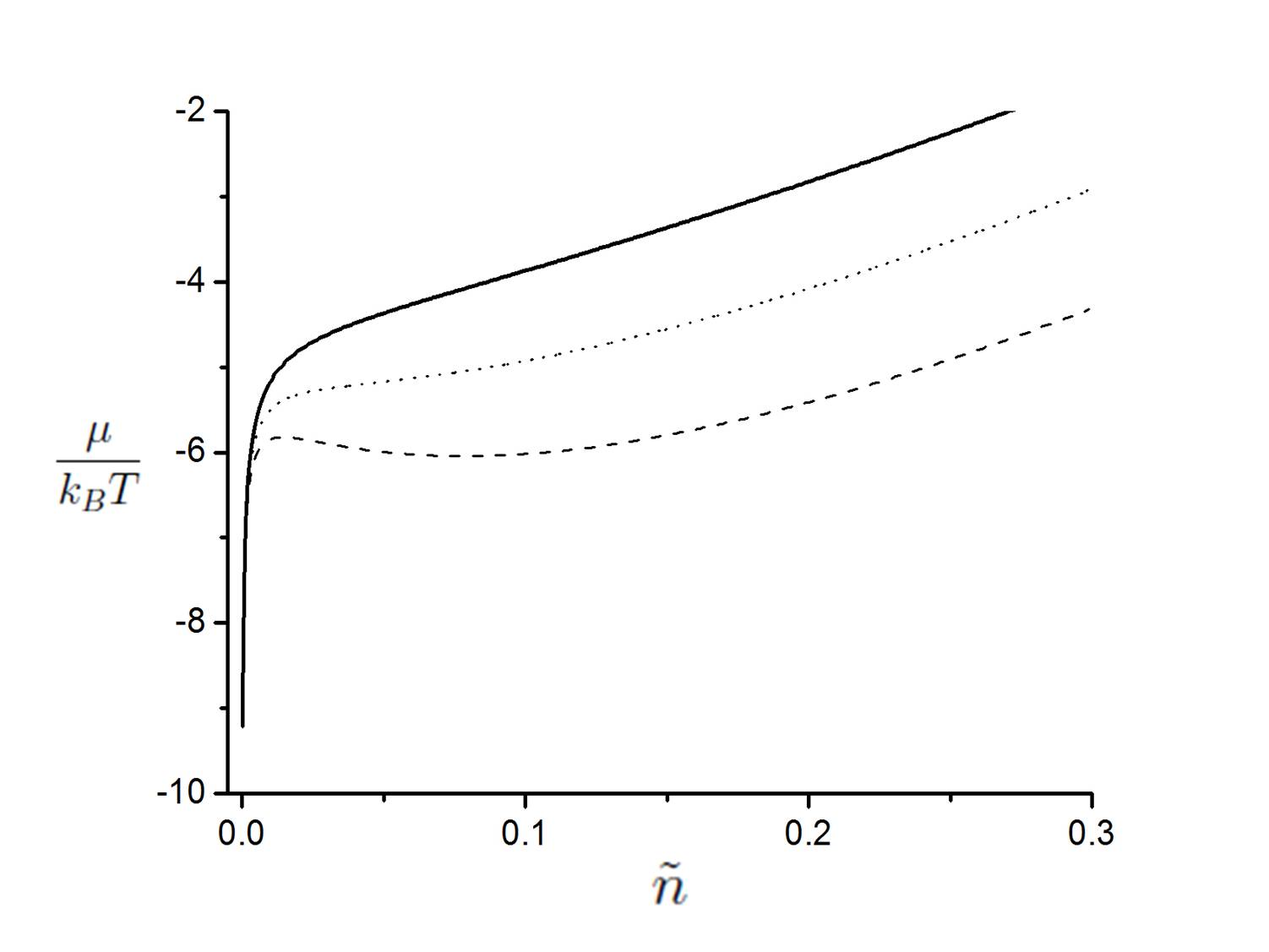}}
\caption{Dependences of chemical potential $\mu=k_B T\ln(nR^3)+\partial{f_{ex}}/\partial{n}$ ($f_{ex}$ is the excess free energy density) on the reduced concentration $\tilde{n}$ at different reduced temperatures: $\tilde{T}=0.25$ (solid line), $\tilde{T}=0.2$ (dotted line) and $\tilde{T}=0.17$ (dashed line). The data are shown for $\tilde{V}_0=1$ and $\tilde{l}=1$.}
\label{fig1}
\end{figure}

\begin{figure}[h!]
\center{\includegraphics[width=0.8\linewidth]{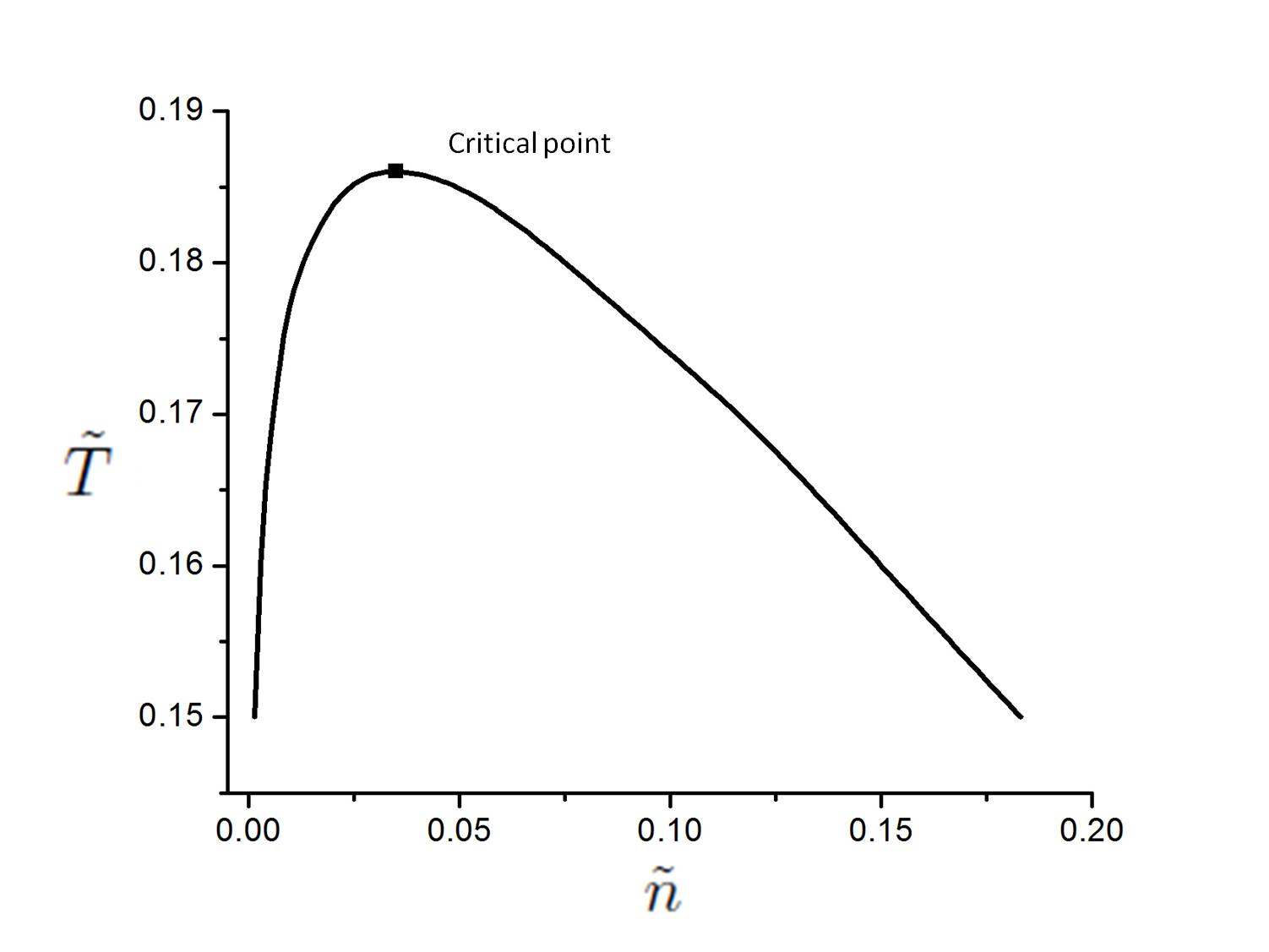}}
\caption{Binodal of the liquid-liquid phase equilibria of a solution of the Gaussian core dipolar particles. The data are shown for $\tilde{V}_0=1$ and $\tilde{l}=1$. The critical parameters for this case are $\tilde{T}_c=0.186$ and $\tilde{n}_c=0.035$.}
\label{fig2}
\end{figure}

Fig. \ref{fig1} demonstrates dependences of the chemical potential, expressed in thermal energy $k_{B}T$ units of the GCDM on the reduced concentration $\tilde{n}$ at different reduced temperatures $\tilde{T}$. As one can see, at a sufficiently high temperature, the chemical potential behaves in a standard way -- monotonically increases with the concentration. However, when the temperature drops below some threshold value, a van der Waals loop occurs on the chemical potential, indicating the liquid-liquid phase separation. Therefore, the present theory predicts that sufficiently strong electrostatic interactions between GCDM in a solution medium can provoke a liquid-liquid phase separation. Fig.\ref{fig2} shows a typical phase equilibrium curve (binodal) for solution of the GCDM with an upper critical point. As is seen, a strongly asymmetric binodal of the GCDM is very similar to the binodal of the restricted primitive model (RPM) of electrolyte solution (see, for instance, \cite{Fisher,Levin}). We would also like to note that the critical parameters for the GCDM have the same order of value that are the critical parameters for the RPM. {\bf We would like to mention that system of Gaussian core dipolar particles with sufficiently large dipolar length is quite similar to systems of soft charged dumbbells which phase behavior also well reproduces the critical behavior of the RPM \cite{Dussi2013,Braun}. The similar critical behavior is realized in the systems of hard charged dumbbells \cite{Daub}.} It is worth noting that quite similar coexistence curves take place in the salt-free polyelectrolyte solutions (see, for instance, \cite{Orkoulas,Budkov2015_2}). It can be suggested that such broad and asymmetric binodals should be realized in the systems where the phase behavior is determined by the long-range Coulomb interactions.

\begin{figure}[h!]
\center{\includegraphics[width=0.8\linewidth]{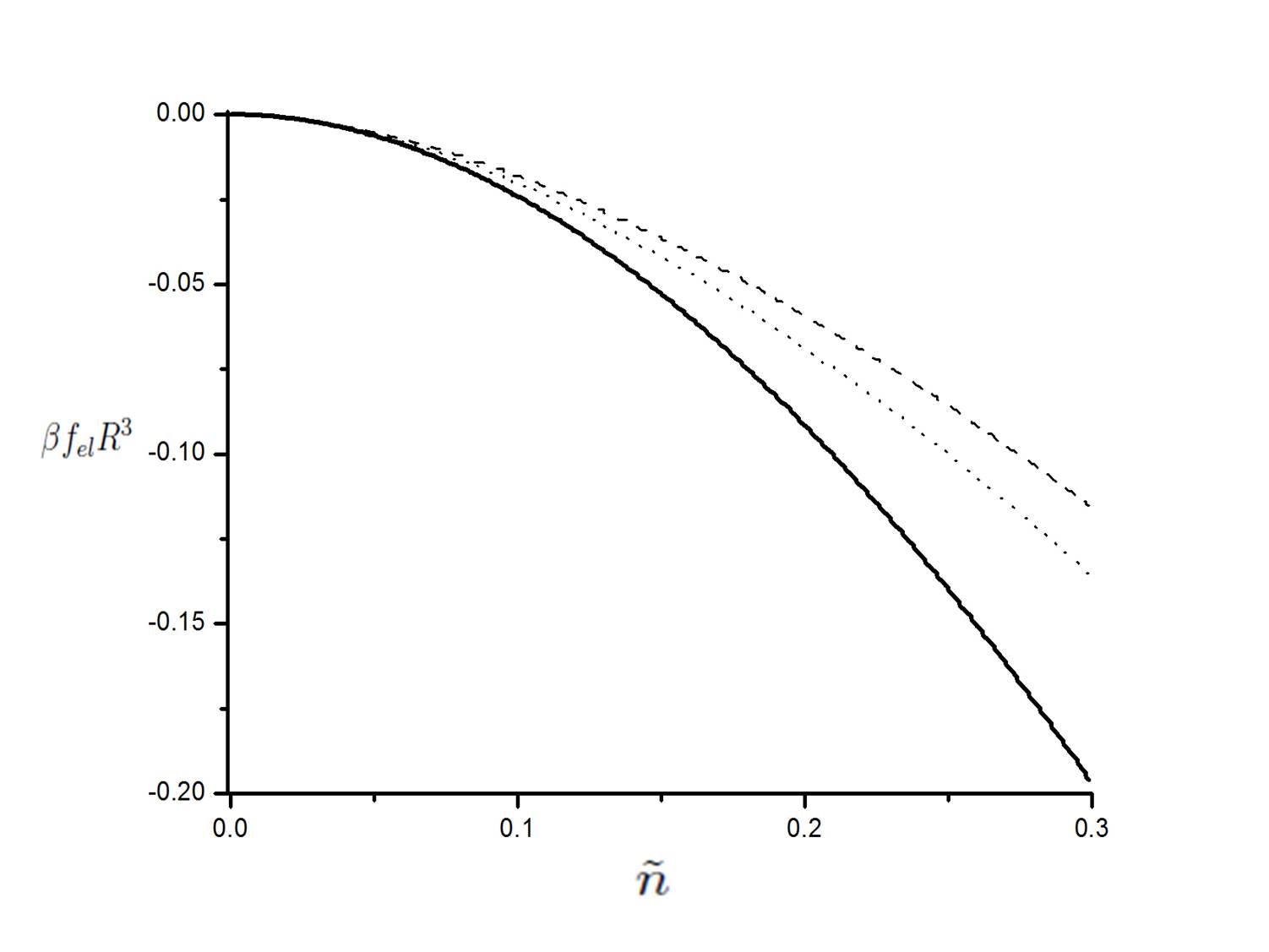}}
\caption{Reduced electrostatic free energy density $\tilde{f}_{el}=\beta f_{el} R^3$ as a function of the reduced concentration $\tilde{n}$ at the fixed temperature $\tilde{T}=1$ and different Gaussian interaction parameters: $\tilde{V}_0=0$ (solid line), $\tilde{V}_0=0.5$ (dotted line), and $\tilde{V}_0=1$ (dashed line). The data are shown for $\tilde{l}=1$.}
\label{fig3}
\end{figure}

We would also like to discuss how the soft core repulsive interactions between the dipolar particles affect the electrostatic free energy of the GCDM. Fig. \ref{fig3} demonstrates dependences of the reduced electrostatic free energy density $\tilde{f}_{el}=\beta f_{el} R^3$ on the concentration $\tilde{n}$ at different parameters $\tilde{V}_0$. As is seen, an increase in parameter $\tilde{V}_0$ leads to a substantial decrease in the absolute value of the electrostatic free energy at rather big concentrations. As it was already pointed out in sec. III, such a behavior of the electrostatic free energy can be easily interpreted. Indeed, making repulsive interaction parameter $V_0$ bigger, we increase the effective excluded volume of dipolar particles, that, in turn, prevents their interpenetration and, consequently, decreases the contribution of the electrostatic interactions to the total free energy.

\section{Concluding remarks and prospects}
We have formulated a field theoretical approach, based on the thermodynamic perturbation theory and random phase approximation for theoretical description of thermodynamic properties of solutions of electrically neutral soft-core molecules with an arbitrary internal electric structure, modelled as a set of charged sites, interacting with each other through the Coulomb and soft-core potentials. We have obtained the general relation for the excess free energy, taking into account internal structure of molecules and electrostatic and soft-core interactions between them. As an illustration, we have applied the developed general theory to theoretical description of a phase behavior of solution of the dipolar particles interacting with each other through the electrostatic and Gaussian potentials -- Gaussian-core dipolar model, which could be considered as a primitive model of dipolar polypeptides or betaines. For the latter model, we have derived an analytical expression for the total excess free energy of solution and, basing on it, described the phase behavior of system. We have predicted that at sufficiently low temperature, when the rather strong electrostatic dipole-dipole attractive interactions take place, the liquid-liquid phase separation occurs. We have shown that the binodal (phase equilibrium curve) has an upper critical point and strongly asymmetric shape. We have established that increase in the energetic parameter of the Gaussian potential at rather large concentrations of the dipolar particles makes the electrostatic free energy less negative.

In conclusion, we would like to discuss further prospects of the developed theoretical background. First of all, our theory can be applied to coarse-grained modelling of the phase behavior of real aqueous solutions of biomacromolecules in salt-free solutions. Secondly, like the previously formulated simpler field-theoretical statistical models \cite{Buyukdagli2013_PRE,Budkov2018}, present theory can be easily extended for the case of presence of salt ions in the bulk solution. The latter will allow us to study an influence of concentration of salt on the discussed in present paper the liquid-liquid phase separation. Thirdly, the proposed theoretical background allows us to formulate the classical density functional theory for solutions of the Gaussian core dipolar particles, extending the density functional theory, proposed in papers \cite{LouisGCM,Archer_1,Archer_2} for the simple Gaussian core model. Such a theory will allow us to describe the concentration profiles of soft-core dipolar molecules at the soft interfaces (with lipid or polymeric membranes), which is relevant for biological and industrial applications. Finally, it is necessary to verify presented here theoretical predictions by means of the Monte Carlo or MD computer simulations. {\bf As it was shown in paper \cite{Braun} by means of MD computer simulation, at sufficiently low temperatures the system of soft charged dumbbells undergoes a liquid-liquid phase separation. It was obtained that critical parameters for soft charged dumbbells are very close to those are realized for the restricted primitive model. Due to the fact that the system of soft charged dumbbells is quite similar to studied here model, one can also expect in the MD simulation the liquid-liquid phase separation for the Gaussian core dipolar model, predicted by present theory.} All these further investigations are the subject of the forthcoming publications.

\begin{acknowledgements}
The publication was prepared within the framework of the Academic Fund Program at the National Research University Higher School of Economics (HSE) in 2019-2021 (grant No 19-01-088) and by the Russian Academic Excellence Project "5-100". The author thanks I.Ya. Erukhimovich for the motivating discussions.
\end{acknowledgements}

\end{document}